\documentclass[aps,prl,twocolumn,groupedaddress,showpacs]{revtex4}
\usepackage{graphicx}
\usepackage{dcolumn}
\usepackage{bm}
\newcommand{\ASJ}{Andreev - Saint-James~}
\newcommand{\YBCO}{YBa$_2$Cu$_3$O$_{7-x}$~}
\begin{document}

\title{Bean-Livingston barrier enhancement on nodal surface of the \emph{d-wave} superconductor YBa$_2$Cu$_3$O$_{7-x}$}


\author{G. Leibovitch} \email{guyguy@post.tau.ac.il}
\author{R. Beck} \thanks{Current address: University of California, Santa Barbara, California 93106}
\author{A. Kohen}
\author{G. Deutscher}
\affiliation{School of Physics and Astronomy, Raymond and Beverly Sackler
Faculty of Exact Sciences, Tel-Aviv University, Tel Aviv, 69978, Israel}


\date{\today}

\begin{abstract}
Vortex entry into (110) oriented YBa$_2$Cu$_3$O$_{7-x}$ films has been studied by tunneling into \ASJ  bound states, whose energy is shifted by surface currents. At low temperatures, the characteristic field for vortex entry has been found to increase up to values several times higher than that of the Bean-Livingston entry field for conventional superconductors, in agreement with recent theoretical predictions.
\end{abstract}


\maketitle

\par
\begin{figure}
{\includegraphics[width=0.7\hsize]{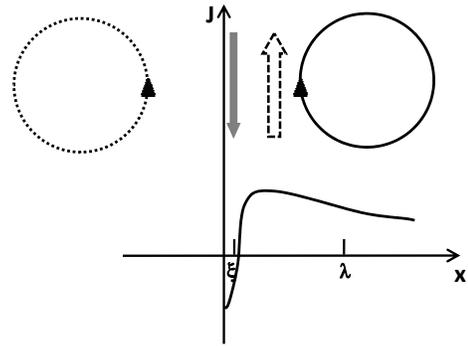}
\caption{Schematic representation of a vortex and its anti-vortex image (dotted) together with the sum of Meissner diamagnetic screening current and the Fogelstr\"om surface paramagnetic current. The vortex anti-vortex interaction is attractive, the effect of the diamagnetic Meissner current is repulsive and that of the Fogelstr\"om  paramagnetic current is attractive. Dashed and full arrows indicate the directions of the paramagnetic and diamagnetic currents. At low temperatures, the sum of both currents becomes paramagnetic on the scale of the coherence length $\xi$. \label{fig_currents}}}
\end{figure}

As shown by C. Bean and J. Livingston \cite{Bean:1968} vortex entry in a Type II superconductor submitted to a magnetic field parallel to its surface can be delayed beyond the field $H_{c1}$ where the mixed state becomes thermodynamically stable in the bulk. Vortex entry may not occur up to a field of the order of the thermodynamical critical field $H_c$ where the free energies of the normal and full Meissner states become equal. The delayed vortex entry comes from the attraction between a vortex and its anti-vortex image. The vortex is however also submitted to a repulsive force from Meissner surface screening currents (Fig. \ref{fig_currents}. At low applied fields these Meissner currents are weak, the vortex - anti-vortex attraction wins and a surface energy barrier prevents vortex penetration in the bulk. At high fields, Meissner currents are strong, the repulsive interaction wins, the surface barrier disappears and vortices can penetrate freely. The attractive and repulsive interactions balance each other at a field of the order of $H_c$. In practice, vortex penetration occurs in general at a field lower than $H_c$, as the sample surface roughness can locally enhance the surface field value above that of the applied field because of demagnetization effects.

\par
\begin{figure}
{\includegraphics[width=0.55\hsize]{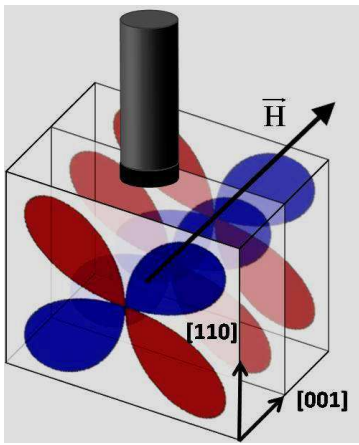}
\caption{Schematic illustration of the experimental setup. The CuO$_2$ planes with the \emph{d}-wave order-parameter are scathed. The thin film surface is [110] oriented. The magnetic field is applied parallel to both the surface and the \emph{c}-axis. The round rod represents the planar tunnel junction (Indium in grey, InO in black).  \label{fig_exp_setup}}}
\end{figure}

Recently, Iniotakis et.al. \cite{iniotakis} have predicted than in a $d_{x^2-y^2}$ wave superconductor having its surface normal parallel to a nodal direction, vortex penetration for fields applied along the \emph{z} direction (the \emph{c}-axis perpendicular to the CuO planes in the cuprates, see Fig. \ref{fig_exp_setup}) can be delayed well beyond the field of first vortex entry $H_{s}$. The increase of the barrier, being particularly significant at low temperatures, is due to is due to strong surface currents carried by \ASJ bound states \cite{FSR}. This current is paramagnetic and flows in the opposite direction with regards to that of the diamagnetic screening Meissner current. A schematic illustration of the currents is presented in Fig. \ref{fig_currents}. One must notice that the Fogelstr\"om paramagnetic current flows at a distance of the order of the coherence length $\xi$ from the surface while the Meissner screening current flows at the scale of the penetration length $\lambda$ from the surface. The effect of the paramagnetic current on vortex entry is to reduce the repulsive force driving the vortex into the bulk and away from the surface.

\par
In this work, we present a study of the temperature-magnetic field dependence of the Bean - Livingston surface barrier for vortex entry on the nodal surface of slightly overdoped YBa$_2$Cu$_3$O$_{7-x}$ (YBCO) thin films. Our results basically confirm the theoretical predictions.

\par
Thin films of YBa$_2$Cu$_3$O$_{7-x}$ (YBCO) were grown by DC
off-axis sputtering on substrates of SrTiO$_3$ with (110)
orientation. A buffer layer of PrBa$_2$Cu$_3$O$_{7-x}$, was
deposited using RF off-axis sputtering on top of the
substrate \cite{Poelders} in order to reduce the (103) orientated
grains. The YBCO growth time was $3$ hours corresponding to a thickness of $2400$\AA. Normal metal - Insulator - Superconductor
(NIS) junctions were prepared by placing thin Indium rods on top of the
surface of the thin film\cite{Krupke,DaganPRL}. Oxidation of the
Indium to InO at the contact area gives the Insulating layer of
the NIS junction. Figure \ref{fig_exp_setup} shows the NIS junction orientations with regards to the \emph{d}-wave order parameter. The samples discussed in this work were all slightly overdoped with a \emph{T$_c$} downset of 89K.

\par
\ASJ zero energy surface bound states (ASJ bound states) are known to develop at surfaces perpendicular to a node direction, such as (110) surfaces in \YBCO \cite{Hu,deutscherRMP}. In a tunneling experiment, these states show up as a Zero Bias Conductance Peak (ZBCP), which splits due to a Doppler effect under superfluid currents flowing parallel to the surface \cite{FSR} as occurs in the geometry show in Fig. \ref{fig_exp_setup}. We have used this split $2\delta$ to investigate the field and temperature dependence of the barrier against vortex entry.

\begin{figure}
{\includegraphics[width=1\hsize]{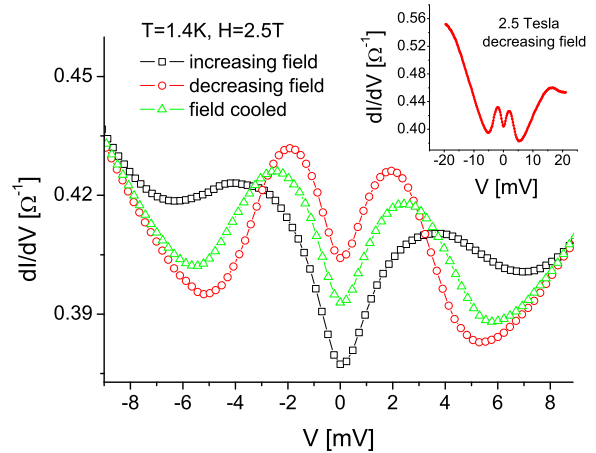}
\caption{$dI/dV$ at 2.5 Tesla for increasing, decreasing and field
cooled conditions. Defining $2\delta$ to be the distance between
the two peaks, we note $\delta$ is maximum for increasing field,
minimum for decreasing field and has an intermediate value for the
field cooled measurement. Inset shows the measurement's full scale
$(\pm20mV)$ for the decreasing field condition. The YBa$_2$Cu$_3$O$_{7-x}$
coherence peaks are located at $\sim16mV$.\label{fig_inc_dec_fc}}}
\end{figure}

In films thicker than the London penetration length as is the case for the samples investigated here, the field evolution is known to be hysteretic, being larger in increasing than in decreasing fields \cite{Krupke, Beck:2004, leibovitch:094522}. The split in increasing fields has been interpreted as due to the Doppler effect induced by Meissner currents \cite{FSR,Covington} while that in decreasing fields has been shown to be a \emph{field} rather than a \emph{current} effect \cite{leibovitch:094522}. A third case is that of a field cooled condition, in which there should be no Bean-Livingston effect. A set of three such measurement is shown in Fig. \ref{fig_inc_dec_fc}. In order to cancel out the \emph{field} effect, we have characterized currents resulting from the Bean-Livingston barrier by the difference between values of $\delta$ measured in increasing field after cooling the sample in zero field and field cooled conditions, ($\delta_{up}-\delta_{F.C.}$). This quantity has been measured as a function of field and annealing temperature $T_a$ as described in the next paragraph.

\par
In the absence of any Bean-Livingston barrier effect, ($\delta_{up}-\delta_{F.C.}$) should reach a maximum value at $H=H_{c_1}$. In the presence of a barrier against vortex entry, $(\delta_{up}-\delta_{F.C.})$ should reach its maximum value at the field $H_S$ where the barrier disappears (following the notations by C. Bean and J. Livingston \cite{Bean:1968}). Tunneling measurement at relative high temperatures have thermal smearing that makes it impossible to follow the ZBCP splitting. In order to follow the temperature dependance of the barrier by tunneling measurements and to eliminate the effect of thermal smearing on the tunneling characteristic, we have pursue the following procedure in determining the temperature dependence $H_s(T)$. After cooling in zero field down to 1.4K, the field is raised up to a value $H^*$. The temperature is then raised up to a value $T_a$ (hence annealing temperature) for a time sufficient to reach equilibrium (2 minutes), then cooled down again to 1.4K for a measurement of $\delta_{up}(H^*,T_a)$. In order to return to the virgin conditions (zero field cooled from above $T_c$), the temperature is then raised up to 105K, the field is lowered to zero, and a new measurement is carried out at a different value of $T_a$. Once several values of $T_a$ have been explored, a new cycle is started with a different value of $H^*$. In order to check if the performed field-thermal cycle damaged the sample or the junction, the tunneling characteristic was measured every time a virgin state was reached and also at the \emph{increasing field} state at the previous $H^*$ value and compared to the tunneling characteristics obtained at the beginning of the measurements. No change in the tunneling characteristic occurred during the measurements.

\begin{figure}
\includegraphics[width=1\hsize]{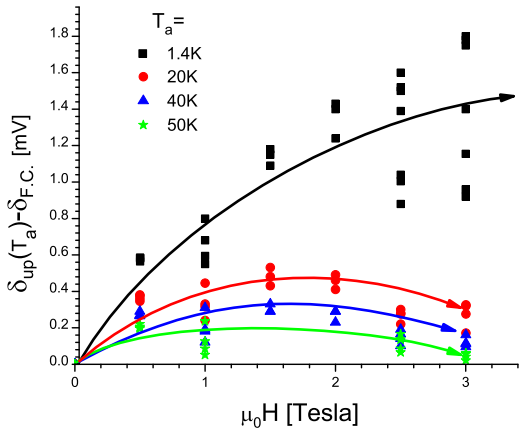}
\caption{The hysteresis in the ZBCP splitting value  $(\delta_{up}(T_a)-\delta_{F.C.})$ for $T_a=1.4K, 20K, 40K$ and $50K$. At $1.4K$ the hysteresis increases until the splitting smears into the background at fields higher then \emph{3 Tesla}. When annealing to higher temperatures, a broad maximum for the hysteresis appears. For $T_a=50K$, the broad maximum is already achieved at field of about \emph{1 Tesla}. Solid arrows are guide to the eye.}
\label{fig_All_Temperatures}
\end{figure}

\par
Fig. \ref{fig_All_Temperatures} summarizes our results for $(\delta_{up}(T_a)-\delta_{F.C.})$ as a function of $H^*$ at different annealing temperatures. At $T_a=20K$ and higher, the curves present a broad maximum at a field $H_M$ that goes down as the annealing temperature goes up, as expected. At the lowest temperatures, no clear maximum can be seen up to the highest field where the value of $\delta_{up}(T_a)$ can be identified.

\par
Ideally, the field dependence of $(\delta_{up}(T_a)-\delta_{F.C.})$ should show a sharper maximum than that which we observe since a reduction in surface currents and hence of $\delta_{up}$ is expected once vortices start to penetrate. Yet, we tentatively identify the field at maximum with $H_S$. What is clear from the data is that surface currents keep increasing up to a field that is much higher than $H_{C_1}$ (of the order of $1\cdot10^{-2} Tesla$) at all temperatures, and than $H_C$ (of the order of $1~Tesla$ \cite{Bean:1968,HcCalc} ) at low temperatures, this basically confirms the prediction of Iniotakis et.al. \cite{iniotakis}. In order to check whether thermal fluctuations are the reason for the continuous change in $\delta_{up}$, a set of measurements was done with much longer time of temperature annealing - up to 6 hours. No difference between the short and long annealing time was found and hence - the thermal fluctuation is ruled out from being a reason for this behavior.

\begin{figure}
\includegraphics[width=1\hsize]{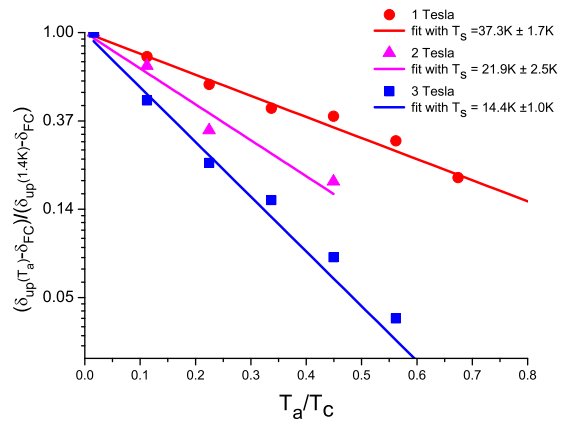}
\caption{The decay of $\delta$ as a function of the annealing temperature $T_a$. The values of $(\delta_{up}(T_a)-\delta_{F.C.})$  are normalized by the value of $(\delta_{up}(1.4K)-\delta_{F.C.})$. Solid Lines are fit to exponential decay of first order. With $T_c=89K$ The fitted order of decay $\frac{T_S}{T_c}$ is $0.16\pm0.01$, $0.25\pm0.03$ and $0.42\pm0.02$ for $1, 2 and 3~Tesla$ respectively.
\label{fig_exp_scale_fit}}
\end{figure}

\begin{figure}
\includegraphics[width=1\hsize]{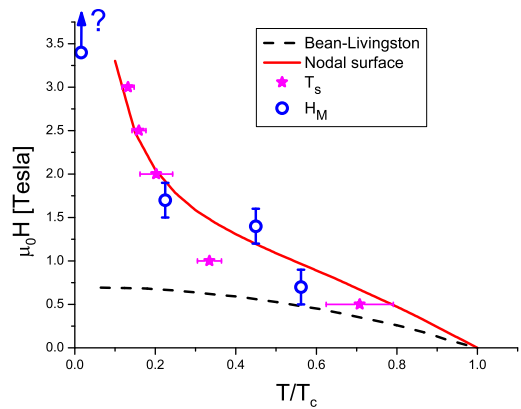}
\caption{Field of first vortex entry as a function of the reduced temperature $\frac{T}{T_c}$. The lines correspond to the Bean-Livingston field of first vortex entry, using $H_c=1~Tesla$ \cite{HcCalc}, in the case of conventional superconductor (dashed line) and in the case of nodal surface calculated by Iniotakis et. al. \cite{iniotakis} (full line).  The circles are represent the field where the hysteresis value of the splitting, $(\delta_{up}(T_a,H^*)-\delta_{F.C.})$ is at maximum that is $H_M$ taken from Fig. \ref{fig_All_Temperatures}. The lowest temperature (noted also with a question mark)is only a lower boundary since no maximum is observed. The stars represent the decay order $T_s(H^*)$ taken from Fig. \ref{fig_exp_scale_fit}. Both methods of analysis show the enhancement of $H_s$ for low temperatures.
\label{fig_final_fit}}
\end{figure}

\par
Another way to represent the results is shown Fig. \ref{fig_exp_scale_fit}, where the dependence of of $\delta_{up}(Ta)-\delta_{up}(1.4K)$ is displayed for a fixed field $H^*$ as a function of $\frac{T_a}{T_c}$. $\delta(H^*)$ is reduced progressively from its value in \emph{increasing field} down to its \emph{field cooled} value as a function of the annealing temperature (\ref{fig_exp_scale_fit}). This reduction can be fitted to an exponential decay, giving a temperature scale $T_s(H^*)$ that decreases as $H^*$ is increased. Instead of this progressive reduction, one would have expected $\delta(H^*)$ to sustain its low temperature value up to the temperature where $H^*$ is equal to the vortex entry field.

\par
There are several possible reasons for the observed broad maximum (Fig. \ref{fig_All_Temperatures}) and for the continuous decay of $\delta_{up}$ (Fig. \ref{fig_exp_scale_fit}). A first one is the surface roughness of the sample that can have weak spots for vortex entry where the surface is not (110) oriented. A second one is the height of the barrier which may become of order $K_BT$ at high fields. One must also take into account the finite thickness of the sample which is only a few penetration depths.

\par
The formation of an $id_{xy}$ subdominant order parameter \cite{leibovitch:094522,galgal}, can be expected to reduce the paramagnetic currents and hence $H_s$. However, at fields of less than 3 Tesla the $id_{xy}$ component disappears above about 7K \cite{galgal}. Therefore, it can not effect the results presented in Fig. \ref{fig_All_Temperatures} and Fig. \ref{fig_exp_scale_fit} at temperatures above 14K. As for the 1.4K  data (Fig. \ref{fig_All_Temperatures}), it only allows to determine a minimum value for $H_s$ which was not calculated by Iniotakis et.al. \cite{iniotakis}.

\par
To summarize the results, we show Fig.\ref{fig_final_fit} values of $\frac{T_s(H^*)}{T_c}$ and of $H_M(\frac{T_a}{T_c})$. Within error bars, both sets of values fall on the same line, showing that both capture the same physics. The dashed line in Fig. \ref{fig_final_fit} is the temperature dependence of the Bean-Livingston field of first entry in a conventional superconductor, taking $H_c=1~Tesla$ \cite{HcCalc}. The solid line is the temperature dependence of the Bean-Livingston field of first entry in the case of a nodal surface, taken from Iniotakis et. al. \cite{iniotakis}, again with $H_c=1~Tesla$. The predicted large enhancement is clearly seen experimentally at low temperatures.

\par
The reason for the low temperature enhancement of the field of first entry is easily understood. As shown by Fogelstr\"om et.al. \cite{FSR} the \ASJ~ states carry paramagnetic currents at nodal surfaces. Iniotakis et.al. \cite{iniotakis} have shown that these currents strongly increase at low temperatures, thus reducing the usual repulsive Lorenz force due to Meissner currents. This reduction of the repulsive force allows the attractive force of the image vortex to remain dominant up to higher fields at low temperatures.

\par
In order to better understand the barrier further theoretical and experimental work is needed in lower temperatures regime where the barrier is strongly enhanced. The doping dependence of the enhancement and especially the appearance of a subdominant nodal order parameter can strongly effect the Bean-Livingston barrier enhancement due to the node removal effect. In such a case, strong over-doped samples and/or high magnetic fields are necessary in order to allow the existence of the subdominant order paramter at temperatures as high as $0.3 T_c$ \cite{leibovitch:094522,galgal}.

\begin{acknowledgments}
The Authors would like to thank S. Hacohen and B. Alomg for the use of their low noise dI/dV measurement system. This work was supported by the Heinrich Herz-Minerva Center for High Temperature Superconductivity and by the ISF.
\end{acknowledgments}

\bibliographystyle{apsrev}
\bibliography{BLonNS}
\end{document}